\documentclass[aps,amsmath,amssymb,reprint,superscriptaddress,prb]{revtex4-2}

\makeatletter
\def\maketitle{
\@author@finish
\title@column\titleblock@produce
\suppressfloats[t]}

\usepackage{dblfloatfix}
\usepackage{xcolor}
\usepackage{float}
\usepackage{graphicx}
\usepackage{dcolumn}
\usepackage{bm}
\usepackage{hyperref}

\usepackage{lipsum}
\begin{document}
\preprint{APS/123-QED}

\newcommand{\thetitle}{High-Coherence  Quantum Acoustics with Planar Superconducting Qubits}

\title{\thetitle}

\author{W.J.M. Franse}
\affiliation{Kavli Institute of NanoScience, Delft University of Technology, PO Box 5046, 2600 GA Delft, Netherlands}
\author{C.A. Potts}
\affiliation{Kavli Institute of NanoScience, Delft University of Technology, PO Box 5046, 2600 GA Delft, Netherlands}
\affiliation{Niels Bohr Institute, University of Copenhagen, Blegdamsvej 17, 2100 Copenhagen, Denmark}
\affiliation{Center for Hybrid Quantum Networks (Hy-Q), Niels Bohr Institute, University of Copenhagen, Copenhagen, Denmark}
\author{V.A.S.V. Bittencourt}%
\affiliation{ISIS (UMR 7006), Universit\'{e} de Strasbourg, 67000 Strasbourg, France}
\author{A. Metelmann}
\affiliation{ISIS (UMR 7006), Universit\'{e} de Strasbourg, 67000 Strasbourg, France}
\affiliation{Institute for Theory of Condensed Matter and Institute for Quantum Materials and Technology, Karlsruhe Institute of Technology, 76131, Karlsruhe, Germany}
\author{G.A. Steele}
\email{g.a.steele@tudelft.nl}
\affiliation{Kavli Institute of NanoScience, Delft University of Technology, PO Box 5046, 2600 GA Delft, Netherlands}

\date{\today}

\begin{abstract}
Quantum acoustics is an emerging platform for hybrid quantum technologies enabling quantum coherent control of mechanical vibrations. High-overtone bulk acoustic resonators (HBARs) represent an attractive mechanical implementation of quantum acoustics due to their potential for exceptionally high mechanical coherence. Here, we demonstrate an implementation of high-coherence HBAR quantum acoustics integrated with a planar superconducting qubit architecture, demonstrating an acoustically-induced-transparency regime of high cooperativity and weak coupling, analogous to the electrically-induced transparency in atomic physics. Demonstrating high-coherence quantum acoustics with planar superconducting devices enables new applications for acoustic resonators in quantum technologies. 
\end{abstract}

\maketitle

In the field of quantum acoustodynamics (cQAD), superconducting transmon qubits \cite{Wallraff_2005,Schuster_2007,Koch_2007} have been coupled to multiple different forms of mechanical modes such as propagating surface acoustic (SAW) \cite{Gustafsson_2014,Satzinger_2018,Bienfait_2019,undershute2024decoherence}, film bulk acoustic resonators (FBARs) \cite{O'Connell_2010}, ultra-high-frequency (UHF) nanoresonator \cite{Rouxinol_2016}, nanomechanical resonators \cite{Arrangoiz_Arriola_2016,Arrangoiz_Arriola_2019} and high-overtone bulk acoustic resonator (HBAR) coupled to either 3D \cite{Chu_2017,Chu_2018,jain2022acoustic} or 2D transmons\cite{Kervinen_2018,Kervinen_2019,Kervinen_2020,Alpo_2021}. The potential applications of HBAR quantum acoustic devices include compact quantum memories \cite{hann2019hardware},  quantum interfaces to spin devices \cite{macquarrie2013mechanical,chen2019engineering}, microwave to optical quantum transduction \cite{blesin2021quantum}, and the exploration of fundamental mass limits in quantum mechanics \cite{gely2021superconducting,bild2022schrodinger}.

An attractive feature of HBARs is the small surface-to-volume ratio of the acoustic modes, enabling them to exploit the exceptionally high bulk mechanical quality factor of crystalline materials, with bulk modes whose quality factor can exceed 10$^{10}$ \cite{galliou2013extremely}. While the pioneering work \cite{Chu_2017} with HBARs achieved moderate quality factors of 10$^5$, quality factors of 10$^6$ were soon achieved \cite{Chu_2018,von_L_pke_2022} by shaping the top piezo element to reduce diffraction by shaping the acoustic mode, creating an effective acoustic plano-convex lens. 

Initial work in the field focused on superconducting qubits in 3D architecture, which enables high coherence times of the superconducting qubit \cite{Chu_2017}. While it can be more difficult to engineer the same coherence times in planar qubit devices, the planar approach offers significant advantages, such as the ease of integrating multiple qubits on the same chip. The planar design also enables the inclusion of fast-flux lines for rapid control of qubit frequencies and qubit couplings and enables high-speed parametric modulation for applications such as parametric amplifiers and on-chip circulators \cite{navarathna2023passive}. The integration of acoustic devices with planar superconducting circuits would provide a valuable new quantum resource, but until now, attempts to integrate HBARs with planar circuits have resulted in reduced coherence of the mechanical modes \cite{Kervinen_2018,Kervinen_2020} exhibiting quality factors on the order of 10$^3$ to 10$^4$.

\begin{figure}[!b]
    \centering
    \includegraphics[width=0.48\textwidth]{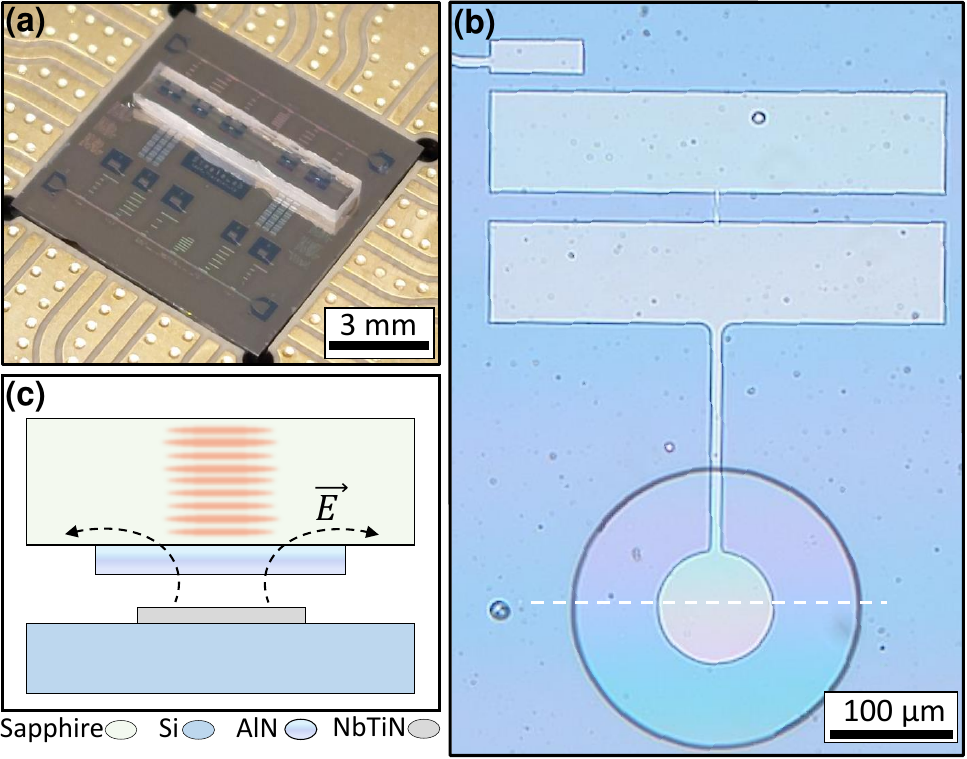}
    \caption{A planar Transmon flip-chip HBAR device. (a) Photo of the assembled device. (b) Optical image of transmon with HBAR on top. White light interference fringes in HBAR suggest a gap distance of a few micrometers between the top and bottom chip. (c) Schematic side view of assembly drawn at the dashed white line in (b). The antenna of the transmon qubit emits an alternating electric field, $\vec{E}$, to actuate the piezoelectric, which in turn generates a standing wave in the sapphire substrate.}
    \label{Figure1}
\end{figure}

Here, we demonstrate the integration of HBAR devices with state-of-the-art mechanical coherence with a planar superconducting qubit chip. Our architecture is based on a flip-chip assembly of a piezoelectric HBAR chip on top of a superconducting qubit chip fabricated on a silicon wafer using standard circuit QED processes and designs. We observe sharp HBAR resonances inside a broad qubit linewidth, whose mechanical nature is confirmed by changing the qubit frequency through multiple resonances via AC-Stark tuning with a nearby drive tone. To decouple the effect of the hybridization of the HBAR mode with the qubit, the resulting lineshapes are analyzed using a master equation simulation to extract the intrinsic mechanical damping rate \cite{Potts_2023lasing}. Doing so, we extract intrinsic damping rates $6.8$ kHz, corresponding to Q-factors of $9.0 \times 10^5$, despite the absence of any mechanical lensing engineered in the piezo actuation layer, promising for the integration of HBARs with more complex superconducting circuits for applications in quantum technologies.

The device has two feedlines with five multiplexed readout resonators coupled to five niobium-titanium nitride transmon qubits. The top line has the flipped HBAR chip aligned with the transmon qubits, and the bottom line is an exact copy without the HBAR chip for control measurements. In this flip-chip architecture, the top chip consists of five cylindrical-shaped aluminum nitride piezoelectric transducers on a sapphire substrate. The bottom chip includes superconducting transmons and coplanar waveguide readout resonators on a silicon substrate; see Fig.~\ref{Figure1}(a,c). An optical micrograph of a single transmon qubit with an HBAR is shown in Fig.~\ref{Figure1}(b); the piezoelectric cylinder is visible as a large circle exhibiting white light interference. The interference pattern suggests a gap between the piezoelectric disk and the silicon substrate on the order of  1 $\mu$m.

To excite HBAR modes near 6 GHz most efficiently, the piezoelectric layer was designed to have a thickness of $t_{\text{p}} \approx 900$ nm. Since the resonance frequency of the piezoelectric disk is given by $f_{\text{0}} = v_{\text{p}} / 2 t_{\text{p}}$, where $v_{\text{p}}$ is the acoustic velocity in the piezoelectric material ($v_{\text{AlN}} \approx 1.14 \times 10^4$ m/s). The mode spacing between two consecutive resonances, the free spectral range (FSR), of the phononic resonator, is given by $f_{\text{FSR}} = v_{\text{s}} / 2t_{\text{s}}$, where $v_{\text{s}}$ is the acoustic velocity in sapphire ($\sim 1.11 \times 10^4$ m/s,) and $t_{\text{s}}$ the HBAR thickness, with a predicted FSR of 8.538 MHz for a 650 $\mu$m sapphire substrate.

\begin{figure}
    \includegraphics[width=0.475\textwidth]{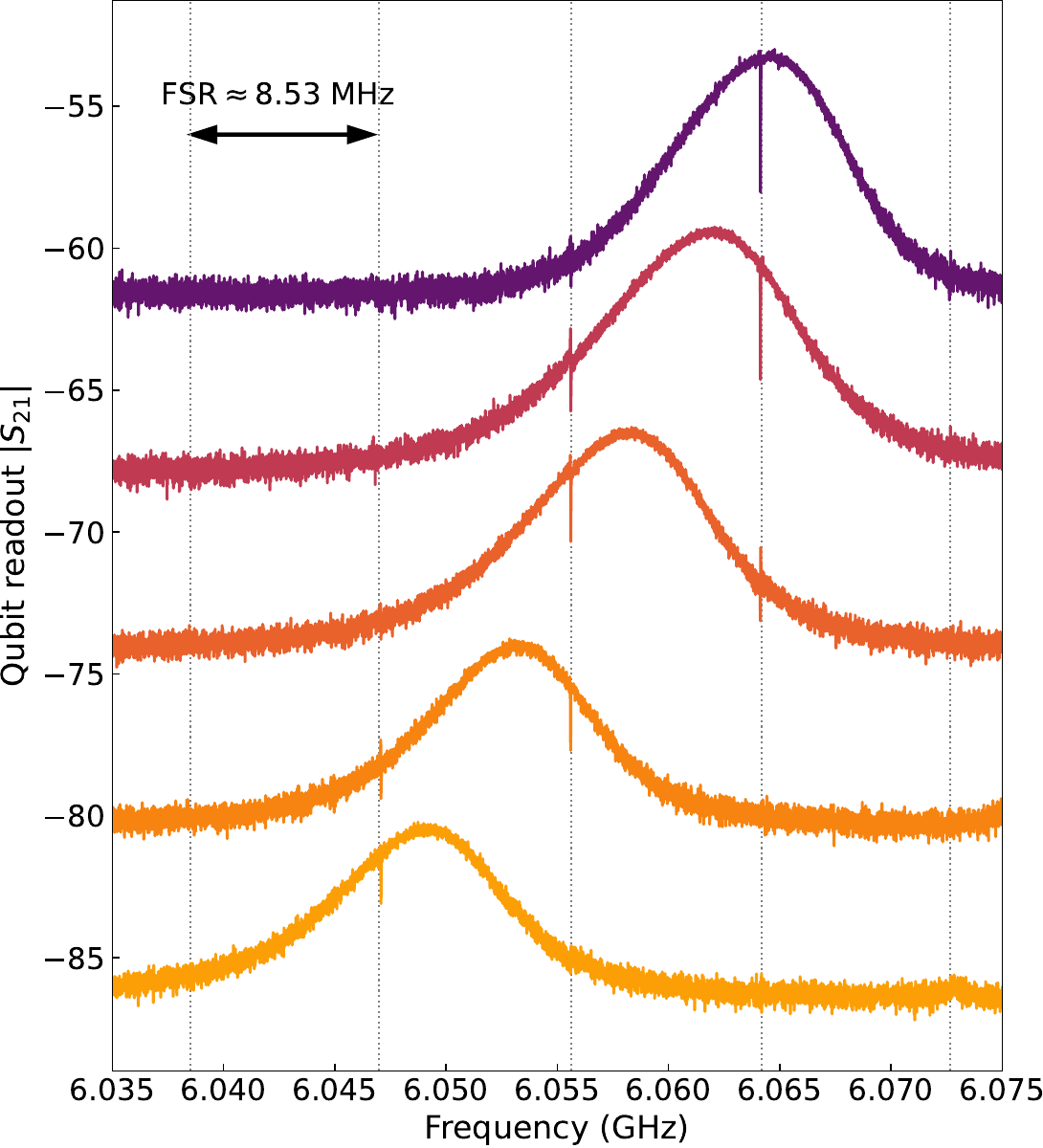}
    \caption{Observation of multiple HBAR resonances by AC-Stark shifting of the qubit frequency.  The qubit $\omega_\text{eg}/ 2 \pi$ transition is tuned using a Stark drive tone with small detuning ($\omega_\text{Stark}/ 2 \pi$ = 6.09605 GHz). The curves represent two-tone spectroscopy measurements for different Stark powers (top to bottom: -22.0, -15.5, -12.0, -9.0, -7.5 dBm). HBAR resonances are observed at fixed frequencies separated by a free spectral range of 8.53 MHz, corresponding to a 650 $\mu$m Sapphire substrate.}
    \label{Figure2}
\end{figure}

The flip-chip device was cooled to a temperature of $\sim$20 mK in a commercial dilution refrigerator. We measured our readout resonator $\omega_\text{r} = 2 \pi \times 4.910$ GHz to be $1.07$ GHz detuned from our qubit's ground-to-excited state transition frequency $\omega_{\rm eg} = 2 \pi \times$ 6.067 GHz. To confirm coupling to HBAR resonances, we use the AC-Stark effect \cite{Brune_1994,Schuster_2005,Gambetta_2006} to shift the qubit's resonance frequency. The Stark drive was applied to the qubit detuned 30 MHz above the qubit frequency, $\omega_\text{Stark} = 2 \pi \times$ 6.09605 GHz. Increasing the Stark tone's power while performing two-tone spectroscopy shifts the qubit to lower frequencies; see Fig.~\ref{Figure2}. In this measurement, a high-power qubit probe tone is applied such that power broadening allows the simultaneous measurement of multiple HBAR resonances within the qubit linewidth. The sharp resonances occur every $8.53$ MHz, in excellent agreement with the designed FSR.

In order to quantify the interaction strength of the qubit and the phonon modes, additional two-tone spectroscopy was performed with sufficient resonator probe power and qubit drive power so that only one acoustic mode became visible within the qubit peak. To optimize measurement time, we performed segmented qubit drive frequency sweeps: In a span of 200 kHz around the acoustic mode, frequency points were spaced 250 Hz apart, but in a larger span of 20 MHz around the qubit $\omega_\text{eg}$ transition, they were spaced 250 kHz apart. The data (grey dots) is shown in Fig.~\ref{Figure3}, where panel (a) shows full acoustically-induced transparency (AIT) of qubit with one coupled acoustic mode and (b) shows a zoom-in of the HBAR mode.

\begin{figure}
    \includegraphics[width=0.48\textwidth]{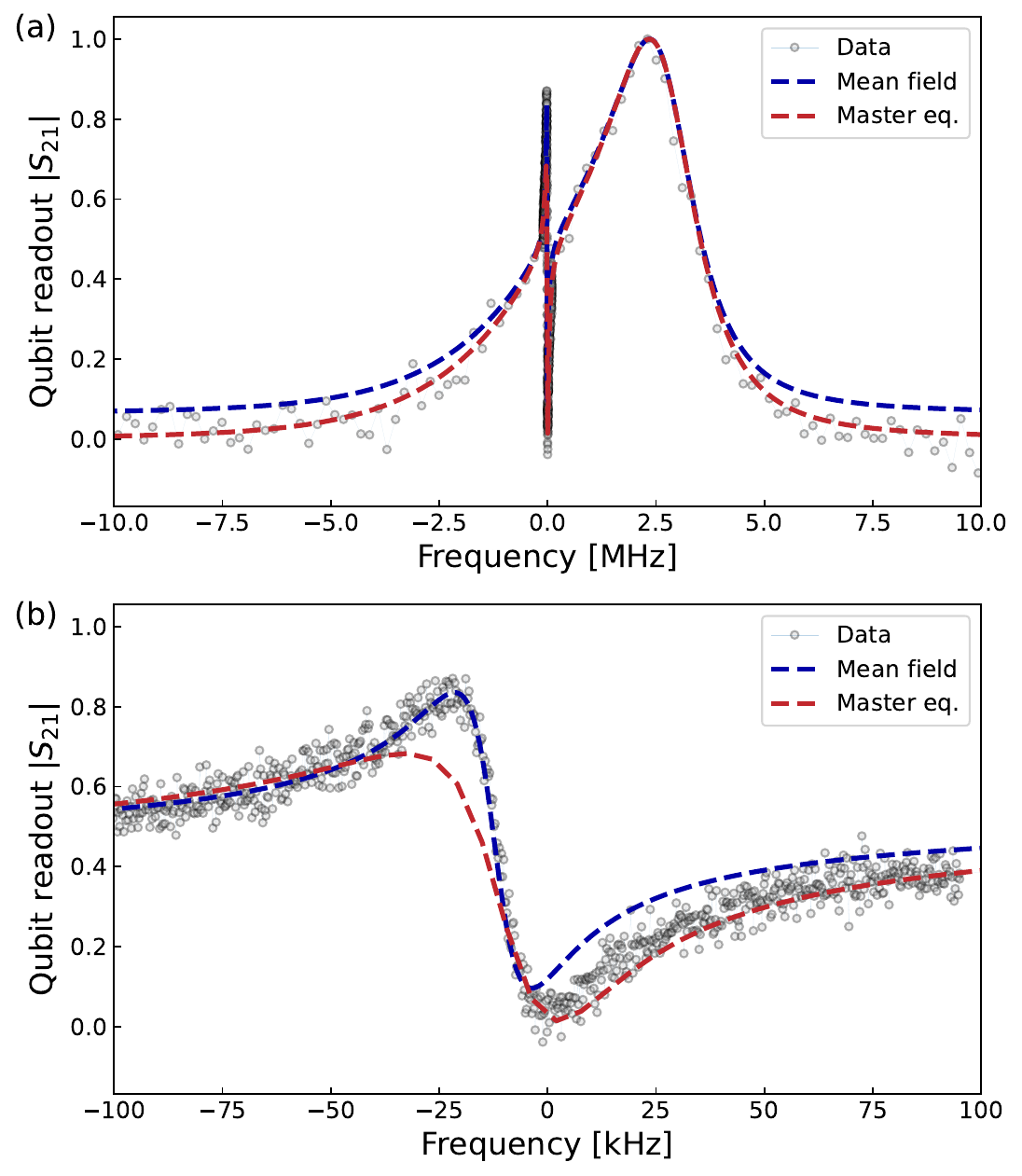}
    \caption{Extracting the intrinsic mechanical damping rate in the AIT regime. (a) Experimental two-tone data of qubit ground-to-excited state transition coupled to a single HBAR mode at 6.064 GHz (grey dots) at low qubit driving power. Both panels also show the qubit spectrum determined from mean-field (blue dashed line) and master equation simulations (red dashed line). (b) Zoom in on the HBAR resonance. The fit to the mean-field theory determines an intrinsic mechanical damping rate of $\gamma/ 2 \pi  = 6.98 \pm 0.03$ kHz, corresponding to a quality factor of 9.0 $\times 10^5$, and a qubit-phonon coupling $g_{\rm{qh}}/2 \pi = 197 \pm 1$ kHz. In panels (a) and (b), frequency refers to the detuning from the HBAR mode.}
    \label{Figure3}
\end{figure}

Figure~\ref{Figure3} shows high-resolution traces of the HBAR AIT resonances observed in qubit spectroscopy taken at low qubit drive and readout resonator drive powers to minimize the broadening of the qubit linewidth. Panel (a) shows a wide sweep in this regime, where the full qubit peak is visible. The qubit spectroscopic peak shows a small degree of asymmetry in its lineshape, which we attribute to the residual thermal occupation of the readout resonator; see Ref.~\cite{Potts_2023lasing}. The line indicates the result of a master equation and mean-field calculations of the qubit spectroscopy. The mean-field curve was used to extract the intrinsic damping rate of the mechanical mode by fitting the data. We note that even at low powers, the qubit has a relatively large linewidth of 425 kHz. While it is likely that dielectric losses from the piezoelectric can reduce qubit coherence, in our case, the reference qubits on the chip which were not coupled to the HBARs also showed comparable linewidths, suggesting the piezoelectric layer, in this case, is not the source of qubit decay. We currently believe that the large linewidth in our qubit fabrication process is related to an interaction of the aluminum layer of the qubit with a silicon layer that was overetched during the base layer patterning. 

Figure 3(b) shows a zoom of the AIT feature with high spectral resolution along with the calculations. From the mean-field model, we extract an intrinsic mechanical damping rate of the HBAR mode of $6.98\pm 0.03$ kHz, describing the mechanical decay rate if the HBAR were decoupled from the qubit entirely, and a qubit-HBAR coupling rate of $197 \pm 1$ kHz. We note that the spectroscopic AIT feature observed in the experimental spectroscopy can be both broader than the intrinsic mechanical linewidth due to hybridization with the qubit but also narrower than the intrinsic mechanical linewidth at higher qubit drive powers due to mechanical amplification and narrowing stemming from single-atom lasing physics \cite{Potts_2023lasing}. As described in previous work \cite{Potts_2023lasing}, there is good agreement between the calculation and the experiment. However, we can see here that the master equation simulations performed failed to capture a slight bump on the left side of the AIT feature. This feature can be captured better by mean-field calculations \cite{Potts_2023lasing}. In any case, both mean-field and master equation modelling describes the behaviour of the data and can be used to extract similar line widths.

Interestingly, we achieved state-of-the-art mechanical coherence, comparable to the best HBARs in 3D qubit architectures, in the absence of any shaping of the piezo surface to produce acoustic lensing. It is interesting to ask why the etched piezo shaping is not needed on our device. One possibility is that the small qubit electrode combined with the large piezo disc may already provide some electrostatic lensing from the non-uniform electric fields in the piezoelectric material, something which could be explored in future work. The high coherence demonstrated here, in any case, opens up an exciting route to combining quantum acoustics with planar technologies such as fast flux lines, flux-mediated parametric gates \cite{didier2019ac}, SNAILs \cite{sivak2019kerr}, and asymmetrically threaded SQUIDs \cite{lescanne2020exponential}.

\vspace{10pt}
W.J.M.F and G.A.S. acknowledge support through the QUAKE project,  project number 680.92.18.04, of the research programme Natuurkunde Vrije Programma's of the Dutch Research Council (NWO). C.A.P. acknowledges the support of the Natural Sciences and Engineering Research Council of Canada (NSERC). A.M. and V.A.S.V.B. acknowledge financial support from the Contrat Triennal 2021-2023 Strasbourg Capitale Europeenne. 

\textbf{Authors contributions} W.J.M.F. fabricated the device, performed experiments, and wrote the manuscript. V.A.S.V.B. performed theoretical modelling. C.A.P. performed experiments, theoretical modelling and provided supervision. A.M. provided supervision and funding acquisition. G.A.S. provided supervision, conceptualization and funding acquisition. 

\textbf{Competing interests}: The authors declare no competing interests. 

\textbf{Data and materials availability} All data, analysis code, and measurement software are available and provided at \url{https://doi.org/10.5281/zenodo.13881198} \cite{Zenodo}.

\newpage
\bibliography{apssamp}

\clearpage
\onecolumngrid 

\setcounter{equation}{0}
\setcounter{figure}{0}
\setcounter{table}{0}
\renewcommand{\theequation}{S\arabic{equation}}
\renewcommand{\thefigure}{S\arabic{figure}}
\renewcommand{\thetable}{S\arabic{table}}

\begin{center}
\textbf{\large Supplemental Materials: High-Coherence  Quantum Acoustics with Planar Superconducting Qubits}
\end{center}

\author{W.J.M. Franse}
\affiliation{Kavli Institute of NanoScience, Delft University of Technology, PO Box 5046, 2600 GA Delft, Netherlands}
\author{C.A. Potts}
\affiliation{Kavli Institute of NanoScience, Delft University of Technology, PO Box 5046, 2600 GA Delft, Netherlands}
\affiliation{Niels Bohr Institute, University of Copenhagen, Blegdamsvej 17, 2100 Copenhagen, Denmark}
\affiliation{Center for Hybrid Quantum Networks (Hy-Q), Niels Bohr Institute, University of Copenhagen, Copenhagen, Denmark}
\author{V.A.S.V. Bittencourt}%
\affiliation{ISIS (UMR 7006), Universit\'{e} de Strasbourg, 67000 Strasbourg, France}
\author{A. Metelmann}
\affiliation{ISIS (UMR 7006), Universit\'{e} de Strasbourg, 67000 Strasbourg, France}
\affiliation{Institute for Theory of Condensed Matter, Karlsruhe Institute of Technology, 76131 Karlsruhe, Germany}
\author{G.A. Steele}
\affiliation{Kavli Institute of NanoScience, Delft University of Technology, PO Box 5046, 2600 GA Delft, Netherlands}

\date{\today}

\maketitle

\begin{figure}[b]
    \includegraphics[width=0.48\textwidth]{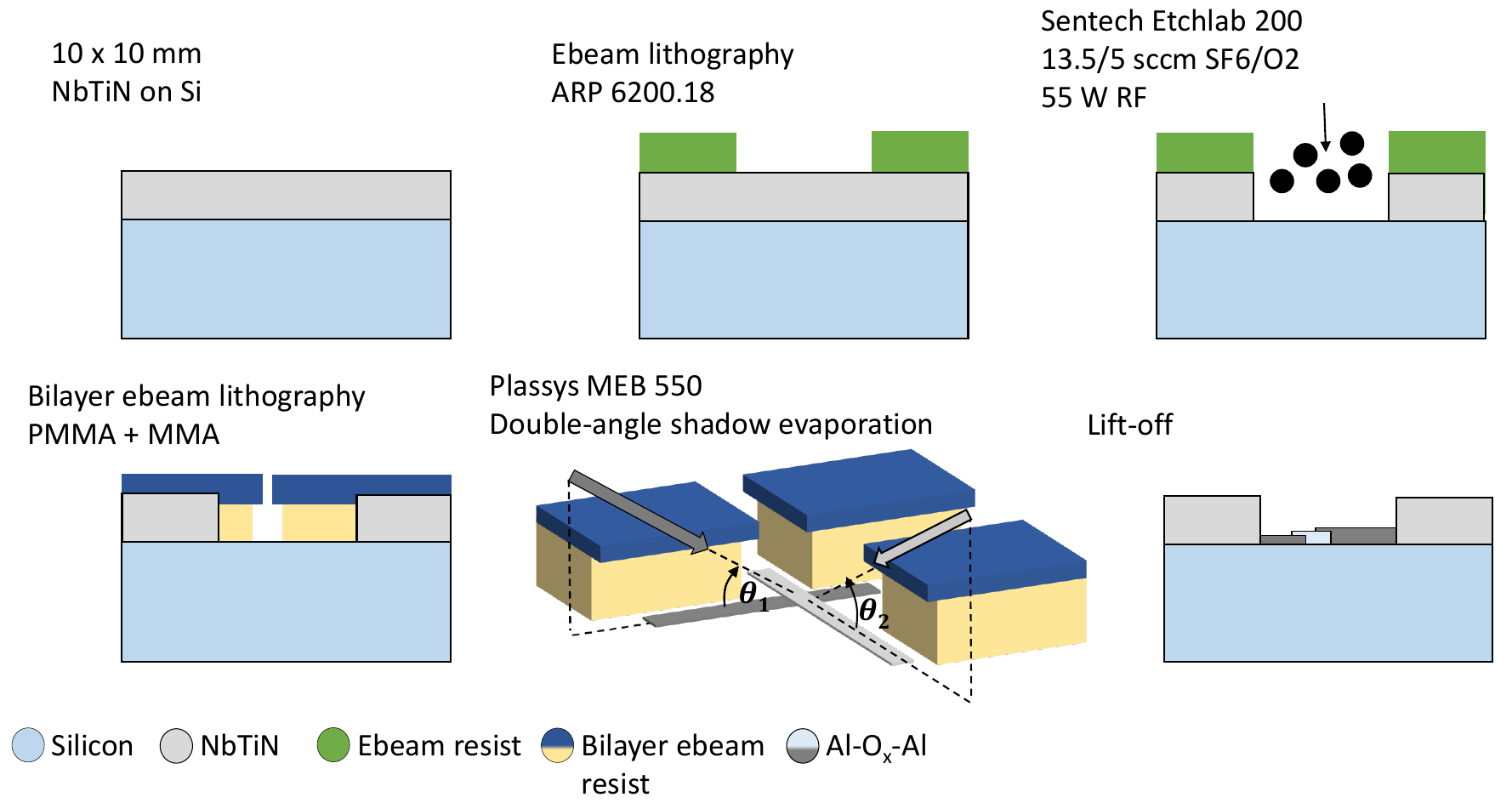}
    \caption{\textbf{Fabrication transmon chip.} Drawings show a step-by-step process for fabricating our 2D transmon qubits. Not to scale.}
    \label{Fig_Qubit_Fab}
\end{figure}

\section{Device Fabriction} \label{Dev_Fab}

\noindent The transmon qubits were fabricated on a 10x10 mm 525 $\mu$m thick high resistivity $\langle100\rangle$ silicon chip with a 100 nm layer of niobium-titanium nitride (NbTiN). The NbTiN deposition step was performed by the Dutch Institute for Space Research (SRON) following the process described in Ref.~\cite{Thoen_2017}. A layer of photoresist (AR-P 6200.18, 4000 rpm) was exposed (EBPG 5200, 315 $\mu$m/cm$^2$) and developed (Pentylacetate, O-xylene, IPA) to form the bulk circuitry (transmon islands and coplanar waveguides). After patterning, the NbTiN was removed using a reactive ion etch (RIE) in the Sentech Etchlab 200 (13.5 sccm $\text{SF}_{\text{6}}$ + 5 sccm $\text{O}_{\text{2}}$, 55 W, 10 $\mu$bar) followed by an in-situ oxygen descum (50 sccm $\text{O}_{\text{2}}$,100 W, 10 $\mu$bar). Following removal of the photoresist layer, a bilayer resist stack (MMA-MAA copolymer 8.5\% EL6, 2000 rpm and P(MMA-MAA) copolymer A6 950k, 1500 rpm, baked for three and five minutes, respectively, at 180 C$^{\circ}$) was used for patterning the 190 nm wide Josephson junctions. The junctions were patterned by e-beam lithography using a dose of 1700 $\mu$C$\text{cm}^{-2}$. The bilayer was developed using cold $\text{H}_{2}\text{O}$:IPA (1:3) and cleaned with IPA. After cleaning the exposed silicon surface with an oxygen descum (200 sccm, 100 W) and an acid clean (BoE(7:1):$\text{H}_{2}\text{O}$, 1:1), the chip was placed in an aluminum (Al) evaporator (Plassys MEB550). The Manhattan-style junctions were fabricated using a standard double-angle shadow evaporation technique with an intermediate in-situ oxidation. The aluminium was deposited at an angle of $35^{\circ}$ at $0^{\circ}$ and $90^{\circ}$ rotation. The bottom and top layers were 35 and 75 nm thick, respectively. After the first evaporation step, the electrodes were oxidized to create the Al$\text{O}_{\text{x}}$ tunnel barriers and following the second deposition step to cap the junctions with a passivation layer. Following the liftoff of the bilayer resist in NMP, the qubit chip was completed. Fig.~\ref{Fig_Qubit_Fab} above shows the step-by-step process just described. 

\begin{figure}[b]
    \includegraphics[width=0.48\textwidth]{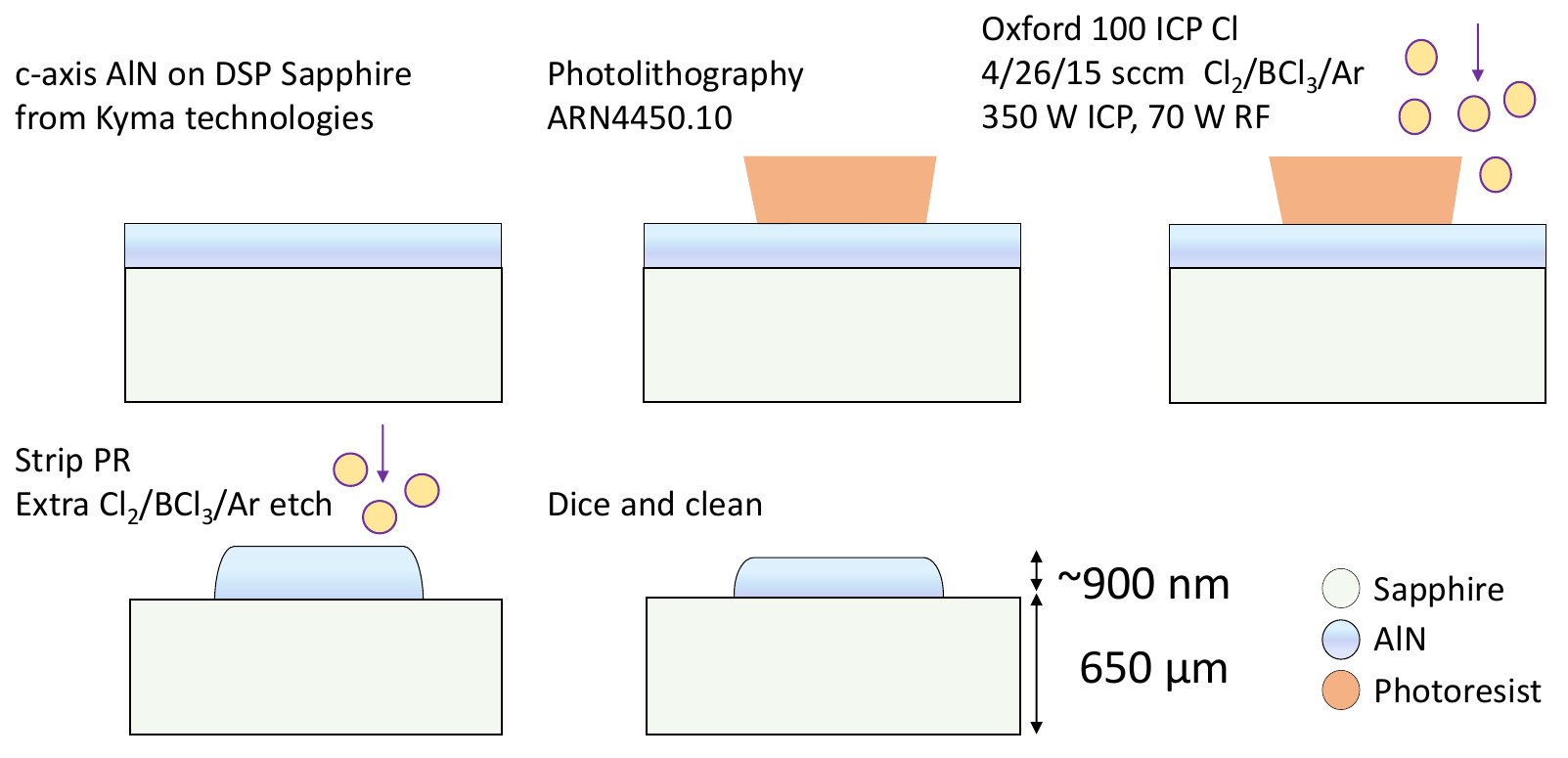}
    \caption{\textbf{Fabrication HBAR chip.} Drawings show a step-by-step process for fabricating our HBAR chip. Not to scale.}
    \label{Fig_HBAR_Fab}
\end{figure}

The HBAR chip was fabricated using double-side polished 4” sapphire wafers; one side was coated with a 1 $\mu$m thick film of c-axis oriented AlN (Kyma technologies, AT.U.100.1000.B). The wafer was first diced into 10x10mm chips for easier processing. A photoresist layer (AR-N 4450.10, 6000 rpm) was used to pattern circular regions to mask off the AlN. The AlN disks were formed using a reactive ion etch (RIE) in an Oxford 100 etcher ($\text{Cl}_{2}\text{/BCl}_{3}\text{/Ar}$ at 4.0/26.0/10.0 sccm, 350 W ICP power, 70 W RF power). After stripping the photoresist, an additional etch step was performed to reduce the thickness of the AlN to $\sim900$ nm. Fig.~\ref{Fig_HBAR_Fab} above shows the step-by-step process just described. 

Once both chips were fabricated, the HBAR chip was diced into 8x2 mm chips. The smaller HBAR chips were flipped on top of the qubit chip (such that the AlN layer facing down). Using probe needles, the AlN disks were aligned with the transmon antennas. Once aligned, the probe needles were used to hold the chips in position, while a tapered fiber was used to apply two-component epoxy (Loctite EA 3430) on the sides of the top chip. After the epoxy dried, the chip was wire-bonded and mounted in a dilution refrigerator. 

\section{Measurement Setup} \label{Setup_extent}
\begin{figure}
    \includegraphics[width=0.48\textwidth]{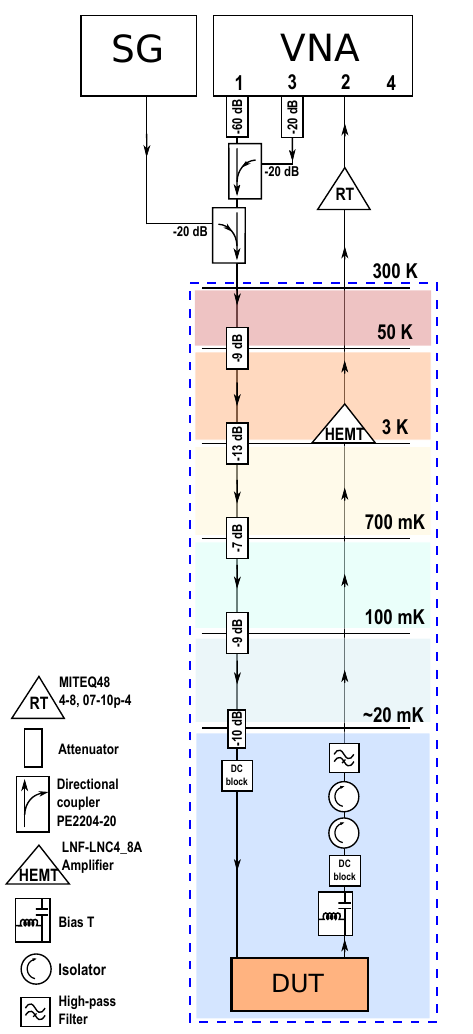}
    \caption{\textbf{Schematic of the measurement setup.} To the right is the general Dilution refrigerator wiring setup. Outside the
    refrigerator, depicted on the top, we used one setup operated for two measurements. In the configuration on the top, we show our two-tone spectroscopy configuration, including an additional signal generator (SG) to perform AC-Stark shifts on our qubit while performing qubit spectroscopy. We sent a weak continuous wave tone (readout resonator probe) from the vector network analyzer (VNA) port 1 and a second continuous wave tone (qubit drive) from the VNA port 3. These two signals are combined using a directional coupler before entering the dilution refrigerator. The signal from the dilution refrigerator goes through a room-temperature amplifier before it goes to port 2 of the VNA. The drive tone coming from the SG is combined with the two VNA signals using a second directional coupler.}
    \label{Schematic_setup}
\end{figure}

\noindent All experiments reported in this article were performed on the baseplate of a commercial dilution refrigerator (Triton200-10 Cryofree dilution refrigerator system, Oxford instruments) operating at a base temperature of $\sim$20 mK. A schematic of the experimental setup and the external configurations used in the different experiments can be seen in Fig.~\ref{Schematic_setup}.

The fabricated sample was wire bonded to a PCB, placed on the dilution refrigerator's mixing plate and connected to two coaxial lines. These coaxial lines were used as input/output microwave lines to measure the readout resonators in a capacitively side-coupled transmission configuration.

\renewcommand{\arraystretch}{1.1}
\setlength{\tabcolsep}{0.5em}
\begin{table*}[t!]
  \begin{center}
    \caption{Device parameters readout, transmon and HBAR.}
    \label{tab:table_device_params}
    \vspace{0.5em}
    \begin{tabular}{p{5cm}|p{2cm}|p{2cm}} 
      Description & 
      Parameter &
      Value \\
      \hline
      Readout resonator frequency \hspace{10em}& 
      $\omega_\text{r}/2 \pi$ & 4.910 GHz \\
      Readout resonator linewidth & 
      $\kappa/2 \pi$ &  2.897 MHz\\
      Resonator Dispersive shift & 
      $\chi/2 \pi$ &  1.2 MHz\\
      Resonator/qubit detuning & 
      $\Delta$ & 1.07 GHz \\
      Resonator detuning & 
      $g_\text{rq}$ & 81.04 MHz\\
      Qubit transition frequency & 
      $\omega_\text{eg}/2 \pi$ & 6.067 GHz \\
      Qubit charging energy & 
      $E_{\text{c}} / h$ &  260 MHz \\
      Qubit Josephson energy & 
      $E_{\text{J}} / h$ & 40.2 GHz \\
      Qubit intrinsic linewidth & 
      $\Gamma/2 \pi$ & 425 kHz \\
      Phonon frequency & 
      $\omega_\text{p}/2 \pi$ & 6.064 GHz \\
      Phonon linewidth & 
      $\gamma/2 \pi$ & 6.98 kHz \\
      Phonon - qubit coupling & 
      $g_\text{qh}/2 \pi$ & 197 kHz\\
    \end{tabular}
  \end{center}
\end{table*}

Outside the dilution refrigerator, coaxial lines connected to a vector network analyzers (VNA, Keysight PNA N5222A, 10MHz-26.5 GHz). Port 1 was employed as a probe signal, port 3 as the qubit drive source and port 2 for readout. For the AC Stark shift experiments, a signal generator was employed (SG, Rohde \& Schwarz (SMB 100A, 100 kHz - 12.75 GHz))

For the probe signal, a total estimated attenuation of -108 dB (excluding cable attenuation) was employed. The Qubit drive included a total estimated attenuation of -148 dB (excluding cable attenuation). The Stark signal had a total estimated attenuation of -68 dB (excluding cable attenuation). The high electron mobility transistor (HEMT, LNF-LNC4\_8A) has a gain of +44 dB. The room temperature amplifier (RT, MITEQ48, 4-8,07-10p-04) has a gain of +35 dB.

\section{Device details} \label{Dev_details}
\noindent Table \ref{tab:table_device_params} gives the device parameter for the readout resonator (RO), transmon qubit and HBAR resonator.

\section{Full master equation simulation} \label{ME_Spect}
In Fig.~3, we show curves corresponding to master equation simulations of the system as well as mean-field modelling. The master equation simulations correspond to numerical solutions of the master equation for the system composed of the resonator, the qubit and the phonon mode, which is given by
\begin{equation}
\label{eq:MEQ}
\begin{aligned}
    \partial_t \rho &= - \frac{i}{\hbar} [\hat{\mathcal{H}}, \rho] + \gamma \mathcal{L}[\hat{b}] \rho +  \kappa \mathcal{L}[\hat{a}] \rho \\
    &+ \Gamma_1 \mathcal{L}[\hat{\sigma}_-] \varrho + \frac{\Gamma_\phi}{2} \mathcal{L}[\hat{\sigma}_{\rm{z}}] \rho,
\end{aligned}
\end{equation}
where $\Gamma_\phi$ is the intrinsic qubit dephasing and $\Gamma_1 \sim 2 \Gamma$ is the intrinsic qubit relaxation given in term of $\Gamma$ the intrinsic qubit linewidth (see table \ref{tab:table_device_params}). The phonon and resonator decay rates are indicated by $\gamma$ and $\kappa$ respectively. The superoperator $\mathcal{L}$ indicates the standard dissipator:
\begin{equation}
\mathcal{L}[\hat{o}]\rho = \hat{o} \rho \hat{o}^\dagger - \frac{1}{2} \{ \hat{o}^\dagger \hat{o}, \rho \}.
\end{equation}
The Hamiltonian $\hat{\mathcal{H}}$ describes a Jaynes-Cummings coupling between the phonon mode and the qubit, and a dispersive coupling between the qubit and the resonator:
\begin{equation}
\begin{aligned}
\frac{\hat{\mathcal{H}}}{\hbar} &= - \Delta_{\rm{b}} \hat{b}^\dagger \hat{b} - \frac{\Delta_{\rm{q}}}{2} \hat{\sigma}_{\rm{z}} - \Delta_{\rm{r}} \hat{a}^\dagger \hat{a} \\
&+ \chi \hat{a}^\dagger \hat{a} \hat{\sigma}_{\rm{z}}+ g_{\rm{qh}}\left( \hat{b} \hat{\sigma}_+ +  \hat{b}^\dagger \hat{\sigma}_- \right) \\
&+ \varepsilon_{\rm{d}} \left(\hat{\sigma}_+ + \hat{\sigma}_- \right)+ \epsilon_{\rm{p}} \left(\hat{a}^\dagger + \hat{a} \right).
\end{aligned}
\end{equation}
The above Hamiltonian is written in a frame rotating at the drive frequency $\omega_{\rm{d}}$. The operators $\hat{\sigma}_{{\rm{z}}}$ and $\hat{\sigma}_{\pm}$ correspond to the standard Pauli operators, while $\hat{a}$ and $\hat{b}$ correspond to the resonator and the phonon mode annihilation operators. The phonon-drive detuning is $\Delta_{\rm{b}} = \omega_{\rm{d}} - \omega_{\rm{p}}$, the resonator-drive detuning is $\Delta_{\rm{r}} = \omega_{\rm{d}}-\omega_{\rm{r}}$ (already including the Lamb shift), and the qubit-drive detuning is $\Delta_{\rm{q}} = \omega_{\rm{d}}-\omega_{\rm{eg}}$. The qubit-resonator dispersive coupling rate is indicated by $\chi$ and the phonon-qubit coupling by $g_{\rm{qh}}$ (see table \ref{tab:table_device_params}). The resonator drive amplitude is $\epsilon_{\rm{p}}$ and the (weak) qubit drive amplitude is $\varepsilon_{\rm{d}}$.

To obtain the red curve shown in Fig.~3 of the main text, we numerically solve the master equation using the Python package qutip and compute the normalized steady-state qubit population as a function of the phonon-drive detuning. The same behavior can be obtained by the procedure outlined in the next section for the mean-field calculations: e.g. by considering the dynamics for an initial density matrix $\hat{\sigma_+} \rho_s$, where $\rho_s$ is the steady-state of \ref{eq:MEQ}, and then evaluating the Fourier transform of the expectation value $\langle \hat{\sigma}_- \rangle (t)$.

\section{Mean-Field Calculation of the qubit spectrum} \label{MF_Spect}

The dynamics of the system can also be described using an effective mean-field approach. The starting point is an effective master equation describing the temporal evolution of the qubit-phonon density matrix $\varrho$, obtained in \cite{Potts_2023lasing}:
\begin{equation}
\label{eq:MEQDEff}
\begin{aligned}
    \partial_t \varrho &= - \frac{i}{\hbar} [\hat{\mathcal{H}}_{\rm eff}, \varrho] + \gamma \mathcal{L}[\hat{b}] \varrho  \\
    &+ \Gamma_1 \mathcal{L}[\hat{\sigma}_-] \varrho + \frac{\tilde{\Gamma}_\phi}{2} \mathcal{L}[\hat{\sigma}_{\rm{z}}] \varrho.
\end{aligned}
\end{equation}
Such a master is obtained by starting with the dynamics of the system in the dispersive cavity-qubit limit, given by the master equation in Eq.~\eqref{eq:MEQ}, and then eliminating the cavity dynamics using a procedure akin to the one in \cite{gambetta2008quantumtrajectory}. The cavity dynamics induces a qubit dephasing and frequency shift, included in the effective Hamiltonian $\hat{\mathcal{H}}_{\rm eff}$ and in the qubit dephasing $\tilde{\Gamma}_\phi$. The effective qubit-phonon Hamiltonian, in a frame rotating at the cavity drive frequency $\omega_{\rm{d}}$, is given by
\begin{equation}
\begin{aligned}
\frac{\hat{\mathcal{H}}_{{\rm{eff}}}}{\hbar} &= - \Delta_{\rm{b}} \hat{b}^\dagger \hat{b} - \frac{\tilde{\Delta}_{\rm{q}}}{2} \hat{\sigma}_{\rm{z}} + g_{\rm{qh}}\left( \hat{b} \hat{\sigma}_+ +  \hat{b}^\dagger \hat{\sigma}_- \right) \\
&+ \varepsilon_{\rm{d}} \left(\hat{\sigma}_+ + \hat{\sigma}_- \right),
\end{aligned}
\end{equation}
The effective dephasing and frequency shift are given by
\begin{equation}
\begin{aligned}
\tilde{\Gamma}_\phi &= \Gamma_\phi + \Gamma_{\phi, \rm{cav}} (t), \\
\tilde{\Delta}_{\rm{q}} &= \Delta_{\rm{q}} - \omega_{\rm{q, cav}}(t),\\
\omega_{\rm{q, cav}}(t)(t) &= 2 \chi {\rm{Re}}[\alpha_{\rm{g}}(t) \alpha_e^*(t)], \\
 \Gamma_{\phi, \rm{cav}} (t) &=2 \chi {\rm{Im}}[\alpha_{\rm{g}}(t) \alpha_{\rm{e}}^*(t)].
\end{aligned}
\end{equation}
In the above expressions, $\alpha_{\rm{g,e}}$ are the microwave cavity occupancies provided that the qubit is in the ground (g) or excited (e) state; they are given by
\begin{equation}
\label{eq:alpe0}
\begin{aligned}
\partial_t \alpha_{{\rm{e}}} &= \left[i (\Delta_{\rm{r}} - \chi) - \frac{\kappa}{2} \right] \alpha_{{\rm{e}}} - i \epsilon_{{\rm{p}}}, \\
\partial_t \alpha_{{\rm{g}}} &= \left[i (\Delta_{\rm{r}} + \chi) - \frac{\kappa}{2} \right] \alpha_{{\rm{g}}} - i \epsilon_{{\rm{p}}},
\end{aligned}
\end{equation}
From the effective master equation \eqref{eq:MEQDEff}, we obtain the following set of equations for the expectation values $\langle \hat{b} \rangle = b,$ $\langle \hat{\sigma}_- \rangle = s_-$, $\langle \hat{\sigma}_{\rm{z}} \rangle = s_{\rm{z}}$:
\begin{equation}
\label{eqs:MF}
\begin{aligned}
\partial_t b &= \left( i \Delta_{\rm{b}} - \frac{\gamma}{2} \right) b(t) - i g_{\rm{qh}} s_-(t), \\
\partial_t s_- &= \left( i \tilde{\Delta}_{\rm{q}} - \tilde{\Gamma} \right) s_- (t) + i g_{\rm{qh}} b(t) s_{\rm{z}} (t), \\
\partial_t s_{\rm{z}} &= 2 i s_-(t) \left( g_{\rm{qh}} b^*(t) + \varepsilon_{\rm{d}} \right) - 2 i s_-^*(t) \left( g_{\rm{qh}} b(t) + \varepsilon_{\rm{d}} \right) \\
&- \Gamma_1( s_{\rm{z}}(t) +1),
\end{aligned}
\end{equation}
which assume the mean field approximations $\langle \hat{b} \hat{\sigma}_{\rm{z}} \rangle \approx \langle \hat{b} \rangle \langle \hat{\sigma}_{\rm{z}} \rangle = b s_{\rm{z}}$ and $\langle \hat{b}^\dagger \hat{\sigma}_{-} \rangle \approx \langle \hat{b}^\dagger \rangle \langle \hat{\sigma}_{-} \rangle = b^* s_{-}$. We have defined the total qubit linewidth $\tilde{\Gamma} = \tilde{\Gamma}_\varphi +\Gamma_1/2$

The qubit spectrum shown in Fig. 3 of the main text is obtained by numerically solving the set of coupled equations \eqref{eqs:MF} with appropriate initial conditions. For that, we recall that the qubit spectrum is given by \cite{Gambetta_2006}
\begin{equation}
\label{eq:qubitabsspect}
\begin{aligned}
S(\omega) &= \frac{1}{2 \pi} \int_{- \infty} ^\infty  dt e^ {i \omega t} \langle \hat{\sigma}_-(t) \hat{\sigma}_+ (0) \rangle_s \\
&= \frac{1}{\pi} {\rm{Re}} \left[  \int_{0} ^\infty  dt e^ {i \omega t} \langle \hat{\sigma}_-(t) \hat{\sigma}_+ (0) \rangle_s \right],
\end{aligned}
\end{equation}
where $\langle \cdot \rangle_s = {\rm{Tr}} [\cdot \varrho_s]$, is the expectation value of an operator in the steady-state $\varrho_s$. We notice then that
\begin{equation}
\langle \hat{\sigma}_-(t) \hat{\sigma}_+ (0) \rangle_s = {\rm{Tr}}[\hat{\sigma}_- \varrho_I(t)],
\end{equation}
where $\varrho_I(t)$ is the temporal evolution of a state initially in $\hat{\sigma}_+ \varrho_s$. Considering the steady-state of the system under a weak qubit drive, which we have studied in detail in \cite{Potts_2023lasing}, we have the following initial conditions for \eqref{eqs:MF}
\begin{equation}
\begin{aligned}
b(0) &= \langle \hat{b} \hat{\sigma}_+ \rho_s \rangle = 0,  \\
s_-(0) &=\langle \hat{\sigma}_- \hat{\sigma}_+ \rho_s \rangle = 1, \\
s_z (0) &=\langle \hat{\sigma}_z \hat{\sigma}_+ \rho_s \rangle = 0. \\
\end{aligned}
\end{equation}
The qubit spectrum is then obtained by the Fourier transform of $s_-(t)$. To obtain the fit parameters given in the main text, we have used standard Python libraries for data fit using the qubit absorption spectrum calculated as explained above.

\section{Reference Qubit Measurements}
\label{Ref_Qubit}

\begin{figure}[b]
    \includegraphics[width=0.48\textwidth]{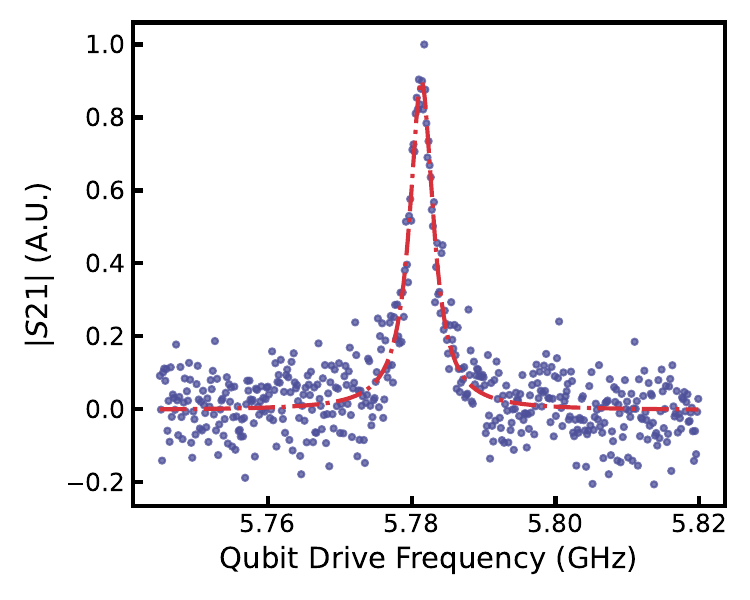}
    \caption{\textbf{Reference qubit two-tone spectroscopy.} Two-tone spectroscopy of a reference qubit.}
    \label{Fig_Ref_Qubit}
\end{figure}

As a reference, a nominally identical qubit was measured without an HBAR to determine the contribution of the qubit decoherence resulting from the coupling to the HBAR. Performing a fit of the qubit spectrum \cite{Gambetta_OurLordAndSavior} yields a qubit linewidth $\Gamma = 2\pi \times$ 1.9 MHz. As a representative qubit, this is quite a bit worse than the value used in the main text, suggesting the coherence limitation of the qubit is indeed unrelated to the piezoelectric material.

Indeed, recent work has identified that the $\text{SF}_{\text{6}}$ etch step is the leading contributor to the qubit decoherence. The original etch has been replaced by a two-step process; the NbTiN was first etched using a $\text{SF}_{\text{6}}$ dry etch through approximately ninety percent of its thickness, leaving a thin layer of metal. The remaining metal was removed using an RCA 1 wet etch; $\text{NH}_3\text{OH}/\text{H}_2\text{O}_2/\text{H}_2\text{O}$ (1:1:5) heated to 40$^\circ$C. This modification to the fabrication recipe has resulted in recent qubits with T$_1$ decoherence times on the order of $30 \mu$s.

\end{document}